\documentclass[twocolumn,showkeys,showpacs,prb,amsmath,amssymb,epsf,aps]{revtex4}
\usepackage{dcolumn}
\usepackage{epsfig}

\usepackage{graphicx}
\usepackage{dcolumn}
\usepackage{bm}

\def\AAA{\AA$^3$}

\def\af{ $\alpha$ }
\def\asos{ $\alpha_{SOS}$}
\def\asp{ $\alpha_{RPA}$}
\def\arpa{ $\alpha_{RPA}$ }
\def\appp{ $\alpha_{PPP}$ }
\def\apen{ $\alpha_{Penn}$ }
\def\aff{ $\alpha_{FF}$}
\def\rm{ $\overline{r}$}

\def\rbar{ $\overline{r}$}

\begin{document}

\title{ Static dielectric response of icosahedral fullerenes from C$_{60}$ to C$_{2160}$ 
 by an all electron density functional theory}

\author{Rajendra R. Zope$^{1}$, Tunna Baruah$^{1}$,  Mark R. Pederson$^{2}$, and B. I. Dunlap$^3$}

\affiliation{$^1$Department of Physics, The University of Texas at El Paso, El Paso, TX 79958, USA }

\affiliation{$^2$Center for Computational Materials Science, Code 6392, US Naval Research Laboratory, Washington, DC 20375, USA}

\affiliation{$^3$Theoretical Chemistry Section, Code 6189,  US Naval Research Laboratory, Washington, DC 20375, USA}

\date{\today }

\begin{abstract}
   The static dielectric response of 
C$_{60}$, C$_{180}$, C$_{240}$, C$_{540}$,
C$_{720}$, C$_{960}$, C$_{1500}$, and C$_{2160}$ fullerenes 
is characterized by an all-electron density-functional method. 
First, the screened polarizabilities of C$_{60}$, C$_{180}$, C$_{240}$, and C$_{540}$,
are determined by the finite-field method using Gaussian basis 
set containing 35 basis functions per atom. 
In the second set of calculations, the unscreened polarizabilities are 
calculated for fullerenes C$_{60}$ through C$_{2160}$ from the self-consistent
Kohn-Sham orbitals and eigen-values using the sum-over-states method.
The approximate screened polarizabilities, obtained by applying  a correction
determined  within linear response theory show excellent agreement with the 
finite-field polarizabilities.
The static dipole polarizability per atom in C$_{2160}$ is ( 4 \AAA\,) 
three times larger than that in C$_{60}$ (  1.344 \AAA). 
Our results reduce the uncertainty in various theoretical models
used previously to describe the dielectric response of 
fullerenes and show that quantum size effects in polarizability
are significantly smaller than previously thought.

\end{abstract}

\maketitle

\section{Introduction}
\label{sec:intro} 

  Reduction in the spatial dimensions in nanoparticles results in 
a number of interesting properties such as, for example, the 
widening of the band gap, reduction in melting temperature,
increased magnetic moments etc.
The fundamental understanding of the size dependence of
properties of nanoparticles, i.e. at what size the quantum 
size effects set in, is important for tailoring these 
systems for possible nanotechnological applications.
Studies along these lines continue to be frontline 
research in nanoscience\cite{haberland:035701,
ramos:045351,wang:123116,yang:201304,
raty:037401,drummond:096801,1996Natur.384..621V,breaux:173401,
vasiliev:1813}.

\begin{figure}
\epsfig{file=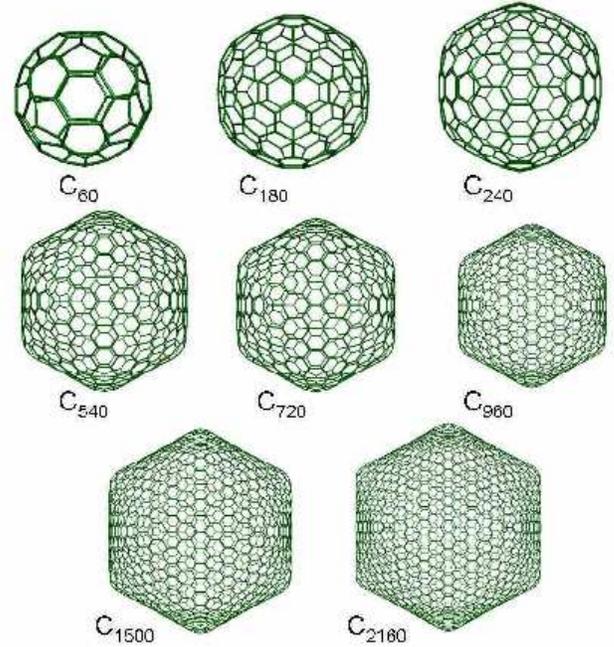,width=8.5cm,clip=true}
\caption{ (Color online) 
The fullerenes structures optimized at ADFT/6-311G** level of theory for which polarizabilities
are calculated. 
\label{fig:str} 
}
\end{figure}

   Fullerenes are hollow cages formed by sp$^2$ bonded carbon atoms. 
 They are finite analogues of graphene
with 12 pentagonal defects that introduce curvature and 
make them closed structures. 
Unlike the most nanoparticles however,
fullerenes are distinctly different in that they have
all the atoms on the surface whereas solid
 spherical nanoparticles have roughly N$^{1/3}$ atoms on the surface. 
 This feature results in 
rapid increase in their volume as the number 
of atoms (i.e. its size) increases.
As dipole polarizability 
is proportional to volume, fullerenes are most suitable 
for investigating quantum size effects manifested in 
polarizability. The large dipole polarizability of carbon 
fullerenes  compared to that of  metallic 
or dielectric spheres of same size have been 
previously interpreted as quantum size effect 
\cite{Pacheco}.
A number of studies on the polarizability of fullerenes
have been devoted to understanding its size 
dependence\cite{C70:376,hu:214708,Pacheco}. 
These studies have been mainly based on classical models 
or have employed drastic approximations to simplify the
polarizability calculations. Unfortunately, the predictions 
of these models are not consistent with each other
and show rather large variance.
For example, the various predictions of the polarizability of  the C$_{2160}$ 
(See Table \ref{table:pol}) are: 
7100 \AAA (Pariser-Parr-Pople model\cite{C70:376}), 
2726 \AAA (bond-order-bond polarizability\cite{hu:214708}), 
9548 \AAA (Penn model-linear response theory\cite{Pacheco}), 
17817 \AAA (tight binding\cite{Pacheco}). 
The predictions from different models of 
fullerene polarizabilities and associated quantum size effects
differ by almost an order of magnitude. 
The most reliable methods to determine electronic 
structure are the first-principles-based quantum-chemical
methods. In the preset work, we use the first-principles 
density-functional methods to accurately determine the 
first order electric response of carbon fullerenes 
of sizes from C$_{60}$ to C$_{2160}$  by computing their static dipole 
polarizabilities. 
These
calculations provide the most accurate prediction 
of the dielectric response of fullerenes in the size 
range C$_{60}$-C$_{2160}$ and show substantial quenching of 
quantum size effects by C$_{960}$ in contrast 
to current understanding.
Moreover, the fullerene polarizabilities computed herein also 
provide an accurate input to the models used in determination of the ultraviolet (UV) spectrum
for possible detection of 
of hyper-fullerene particles  
in space\cite{Astro}.

\section{Theory, Results and Discussion}
  
   The first essential step in obtaining an accurate 
estimate of polarizability of fullerene involves
accurate determination of the fullerene structure. 
We optimized the geometry of carbon fullerenes using recent fully 
analytic and variational formulation of density 
functional (ADFT) 
theory\cite{PhysRevA.52.R3397,werpetinski:7124,RF:204,RF:110,RF:108,RF:194}. 
Being free 
of numerical grids, the ADFT implementation is 
computationally very efficient and allows 
calculations of matrix elements and energies 
accurate to machine precision. Its functional
form is restricted but its space of atomic 
parameters is rich. It has been successfully
used to study electronic properties of 
boron nitrides, aluminum nitride tubes\cite{RF:3,RF:4,RF:119},
and  provides dipole moments, total and atomization 
energies that are comparable to those obtained 
using some of the most sophisticated density functional models.
 In this work, we optimized the geometries of  C$_{60}$,  C$_{240}$, C$_{540}$, C$_{960}$, 
C$_{1500}$, and C$_{2160}$ fullerenes 
from the 60$N^2$ family of icosahedral fullerenes. Additionally, two fullerenes
(C$_{180}$ and C$_{720}$) from the 180$N^2$ family are also optimized.
The structures (Cf. \ref{fig:str})
are optimized at the {\em all }
electron level using the triple-zeta-quality (6-311G**)
orbital basis\cite{mclean:5639} and 
an exchange-correlation basis set that includes
$f$ functions, and the exchange-correlation parametrization 
that gives an exact C$_{60}$ geometry\cite{RF:106}.
These approximations  give accurate geometries in a very  efficient 
manner, which we verified by reoptimizing the larger
C$_{240}$ fullerene using even larger NRLMOL basis 
(35 basis functions per atom) and using the more complex and popular
Perdew-Burke-Ernzerhof (PBE)\cite{RF:183}
generalized gradient approximation (GGA). 
Both models give identical C$_{240}$ geometries
within a root-mean-square value  of 10$^{-3}$ a.u.
for the forces. 




        For the polarizability calculations the  fullerenes structures are 
held rigid 
i.e., the ionic (vibronic) contribution to dipole 
polarizabilities is ignored. This is a tenable 
assumption as our earlier studies on 
C$_{60}$ and C$_{70}$ show that in fullerenes electronic part 
determines the total polarizability with negligible contributions from the ionic vibrations 
\cite{PBAS05,RRZ07}.

\def\AAA{\AA$^3$}
\begin{table*}
 \caption{The average ionic radius \rm (in \AA) and polarizabilities (in \AAA)
obtained in various models: 
     bond-order bond polarizability (BOOP)\cite{hu:214708}
     \appp: Pariser-Parr-Pople approach\cite{C70:376}, 
     \apen: Classical electrostatic within linear response theory (Penn model)\cite{Pacheco},
      TB: tight-binding  \cite{Pacheco},
      \asos: Sum-over-states-ADFT (present),
      \asp: Sum-over-states-ADFT-RPA (present),
      \aff: finite-field PBE-GGA (DFT) (present),
    $\epsilon$: estimates of static dielectric constant for fcc structures.}
\label{table:pol}
\begin{tabular}{lrrrrrrrrrrr}
\hline
   Fullerene &~~~\rm & ~BOPP  & ~~\appp & ~~\apen & ~~TB  &  ~~\asos &  ~~~\asp & ~\aff  &   ~~$~\epsilon$ \\
\hline         
C$_{60}$   & 3.55&    76 &    80 &      64 &      81 &     292 &     79 &    82 &    4.50 \\
C$_{180}$  & 6.13&   227 &   209 &       - &       - &    1202 &    300 &   295 &    4.76 \\
C$_{240}$  & 7.07&   303 &   306 &     343 &     581 &    1754 &    432 &   441 &    4.79 \\
C$_{540}$  & 10.27&  681  &   866 &    1026 &    1869 &    4823 &   1155 &  1193 &   4.71 \\
C$_{720}$  & 12.17&  909 &  1390 &       - &       - &    8000 &   1848 &     - &   4.96 \\
C$_{960}$  & 14.03&  1212 &  2150 &    2185 &    4290 &   12023 &   2745 &     - &   5.00 \\
C$_{1500}$ & 17.94&  1893 &  4415 &       - &       - &   29706 &   5690 &     - &   5.61 \\
C$_{2160}$ & 21.01&  2726 &  7100 &    9548 &   17817 &   38288 &   8548 &     - &   5.01 \\
\hline
\end{tabular}
\end{table*}


  We have determined polarizabilities within two computational 
models.   In the first set of calculations 
polarizabilities of C$_{60}$, C$_{180}$, 
C$_{240}$, and C$_{540}$ are obtained by the {\em finite-field}
method\cite{cohen:s34,Bartlett_FF,Kurtz82,
Salahub_Pol,Na_Pol,Jackson}, using the NRLMOL 
code\cite{RF:69,RF:74,RF:71,RF:184}. In NRLMOL,
optimal numerical grids are used to obtain the
energies and the contributions to matrix elements from 
the exchange-correlation terms.
It is well known that an accurate description of 
electric response requires use of high quality basis 
sets. Here, we use large Gaussian basis sets 
consisting of  5 $s-$, 3 $p-$, and 1 $d-$type Gaussians 
each contracted from 12 primitive functions\cite{RF:181}. 
This basis set is given in Table \ref{tab:basis}.
More details about the basis set can be found in 
Refs.\onlinecite{RF:181,PB_Pol,PBAS05}.
Further inclusion of long range d-type basis functions
increased polarizability of C$_{60}$ by only 0.6\%. Therefore, the 
polarizabilities \af of 
C$_{180}$, C$_{240}$ and C$_{540}$ were computed without 
any additional diffuse functions to avoid
linear dependency of basis sets. 
The total number of basis functions in the finite-field polarizability 
calculations with NRLMOL range from 2100 in C$_{60}$ 
to 18900 in C$_{540}$.
The latter is comparable to the 38,880 orbital
basis functions used in our ADFT calculation on C$_{2160}$.
The mean polarizability is obtained from the
diagonal elements of the polarizability matrix whose 
components can be determined either from the total energies 
or from the induced dipole moments at various field strengths.
The polarizabilities obtained using these two approaches agree within 
1\%.  The finite-field \af values are given in the third-to-last
column of Table I. Although the finite-field method is perhaps the best approach to
calculate polarizabilities, the 
computational cost  becomes prohibitive for
 larger fullerenes. The cost rises principally due to the very large number of 
self-consistent cycles required to obtain tightly converged energies (10$^{-9}$ a.u.) in presence of the 
external electric fields. It is necessary to  keep the convergence criteria tight
to extract meaningful estimates of polarizability
by the finite-field method.  Therefore, we resort to an 
approximate technique that uses the unscreened polarizability obtained 
by the sum-over-states (SOS) method and the linear response theory and random phase approximation
  to estimate the
screened polarizabilities from the $\alpha_{SOS}$.  
In the independent electron model, the working SOS  expression for the $xy$ 
component of polarizability tensor can be written as,
\begin{equation}
\alpha_{xy} = 2 \sum^{unocc.}_{m} \sum^{occ.}_{n}  \frac{<\psi_m|\hat{x}|\psi_n><\psi_n|\hat{y}|\psi_m> }{ \mid 
\epsilon_m - \epsilon_n \mid }.
\label{Eq:SOS}
\end{equation}
Here, the $\{\psi_i\}$ are single particle orbitals and the $\epsilon_i$ is the single particle energy
of the $i^{th}$ orbital.
The $\alpha_{SOS} = \frac{1}{3}(\alpha_{xx} + \alpha_{yy} + \alpha_{zz} ), $
is calculated  according to Eq. (\ref{Eq:SOS}) once the 
self-consistent solution of the Kohn-Sham equations for 
each fullerene is obtained using the ADFT code. 
The computational efforts involved in determining the large 
number of transition dipole moments are reduced by making 
use of group theoretical techniques. 
The calculated  \asos\, for fullerenes C$_{60}$ through C$_{2160}$
are given in the correspondingly labeled column of Table \ref{table:pol}. 
The unscreened \asos\, for C$_{60}$  is  292 \AAA, which compares 
well (282\AAA\cite{Westin} and  311\cite{RF:59}) 
with previous predictions with different 
basis sets and exchange-correlation 
functionals.

The  \asos\, determined using Eq. (\ref{Eq:SOS})  gives unscreened 
polarizability since  the field-induced polarization 
effects (local field effects) are missed in the calculations.
Consequently, the \asos\, overestimates the experimentally 
accessible screened polarizability\cite{Zope:cond-mat0701466}.
For the smallest fullerene (C$_{60}$), \asos\ is roughly 3.6 
times larger than the screened polarizability in 
agreement with previous reports\cite{RF:59,Westin}.
This overestimation slightly increases (to a factor $\sim 4$) 
for C$_{180}$, C$_{240}$, and C$_{540}$ as can been seen from comparison 
of \asos\, and \aff\, from Table \ref{table:pol}.
The screened polarizability can be deduced from \asos\, using  the 
linear response theory \cite{Benedict}
and the random phase approximation (RPA) by means of the
following expression:
\begin{equation}
 \alpha_{RPA} = \Biggl [ 1 + \frac{\alpha_{SOS}}{R^3} \Biggr ]^{-1}
    \alpha_{SOS}.
    \label{eq:RPA}
\end{equation}
Here,
$R$ is the fullerene radius\cite{Benedict}. For fullerenes (larger 
than C$_{60}$) that are polyhedral in structure, the average 
ionic radius can be used.  The average ionic radii \rbar,
are also given in Table ~\ref{table:pol}. Instead of using the ionic radii
of fullerenes in Eq. (\ref{eq:RPA}) which tend to overcorrect the 
screening, it is more appropriate to  use an effective radius
$\overline{r} + \delta$,. The factor $\delta$, accounts for 
the $\pi$ electron cloud around the bare nuclear skeleton. Its typical 
values are 1.2-1.3 \AA\cite{Benedict}. 
Similar methods 
of determining the screened electronic response has 
been used in past for carbon fullerenes\cite{Bertsch,Pacheco}, 
multishell carbon fullerenes- carbon onions\cite{Ruiz}, 
and (cylindrical) carbon nanotubes\cite{Benedict}.  However, 
as we shall see, an accurate initial self-consistent treatment is 
necessary to get a meaningful estimate of the screened polarizability
from Eq. (\ref{eq:RPA}).  
The \arpa\, determined  using  Eq. \ref{eq:RPA} (with $\delta = 1.23$\AA)
are listed in Table \ref{table:pol}. 
It is apparent that \arpa obtained within the ADFT are quite accurate 
as can be gleaned from comparison of \aff\, and \arpa 
for C$_{180}$,  C$_{240}$ and C$_{540}$. 
The \arpa is within 3\% of \aff\, for these 
fullerenes. This excellent agreement  provides confidence in the
\arpa\, values of the larger fullerenes for which a direct
calculation of the self-consistent 
response by means of  the finite-field method is
impractical.  

 The electronic structure of these fullerenes show 
that the highest occupied molecular orbital (HOMO) 
and the lowest unoccupied orbital (LUMO) belong 
to the $h_{u}$ and $t_{1u}$ 
irreducible representations of the icosahedral point 
group. The 
lowest HOMO-LUMO ($h_u \rightarrow t_{1u}$) 
excitation is forbidden by dipole selection rule and does not
contribute to the polarizability.  The inspection 
of the response of smaller fullerenes indicate that the low
energy excitations are strongly screened by application 
of Eq. \ref{eq:RPA}. A detailed study of optical 
spectra will be published elsewhere. 

   In Table \ref{table:pol}, the prediction of various models 
for polarizability are compared. The $\alpha_{Penn}$ is the 
polarizability determined using a semiclassical model that
uses classical electrostatics and linear response theory.
The fullerene polarizabilities larger than the $\alpha_{Penn}$
value have been interpreted as quantum size effects\cite{Pacheco}.
The currently existing understanding, based on the tight-binding 
study\cite{Pacheco}, is that
quantum size effects play very important role in electronic response
of fullerenes up to C$_{3840}$. In contrast, our first principle 
calculations (\aff\, and \asos\, in Table \ref{table:pol}) show 
that by C$_{2160}$ the semiclassical \apen\, is already larger 
than the first-principle value.
For the smaller fullerenes these effects are also much less 
pronounced than  predicted by tight-binding 
model. 
For C$_{240}$, the tight binding value of 
polarizability exceeds finite-field DFT value by 32\%.
Also evident from 
Table \ref{table:pol} is that, while all models well predict
known C$_{60}$ polarizability, they are inadequate to 
provide proper description of response of larger fullerenes.
The tight binding approximation in particular greatly 
exaggerates the response leading to large \af values.
As same Eq. (\ref{eq:RPA}) has been used in tight-binding
study\cite{Pacheco}, it is clear that proper description 
of dielectric response require accurate self-consistent
density functional solution.  The table 
also indicates dramatic increase in polarizability per atom
with fullerene size. The polarizability per atom 
systematically increases from 1.34 \AAA\,
in C$_{60}$ to 4 \AAA\, in C$_{2160}$.  There is no obvious reason
why the polarizability per atom will not increase indefinitely as 
the rigid structure of fullerenes will not allow collapse.
Furthermore, the fullerenes become energetically more stable with
increasing size.  The only problem is physically making them.

\begin{figure}
\epsfig{file=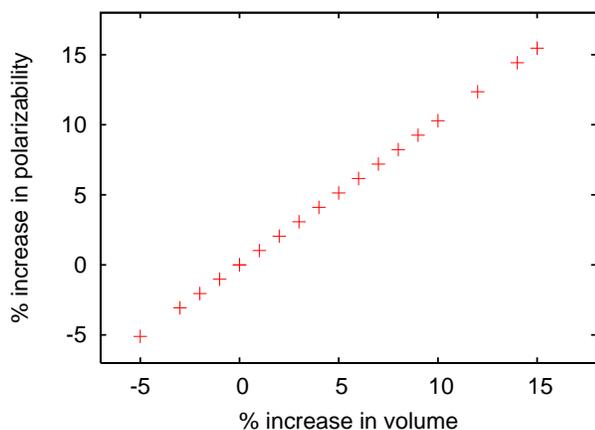,width=8.5cm,clip=true}
\caption{ (Color online) The percent 
increase in polarizability, $\alpha_{RPA}$, 
as a function of percent increase in volume, $\overline{r}^3$,
for C$_{960}$ fullerene.
\label{fig:scale}
}
\end{figure}

 The \arpa\, values given in Table I are the most accurate predictions
of fullerene polarizabilities. The quantitative agreement with 
experimental values of polarizabilities require a careful treatment 
of all possible contributions to polarizability. The vibrational 
contributions as noted earlier are insignificant\cite{PBAS05}.
The excellent agreement between the predicted polarizability and experimental 
measurements for C$_{60}$ and C$_{70}$ rule out the role of 
temperature effects\cite{PBAS05,RRZ07}. 
Previous studies on carbon nanostructures have indicated negative thermal (volume)
expansion at low temperature\cite{Musfeldt,PhysRevB.68.035425,Japan}.
We estimate the effect of volume expansion
by calculating \arpa\, for all fullerenes within 15\% of their equilibrium 
volume. The  $\alpha$ increases practically linearly with increase 
in fullerene volume within the chosen range of volume 
expansion (Cf. Fig. \ref{fig:scale}). This, as expected, 
highlights the importance of accurate geometries for accurate 
prediction of fullerene polarizability.

 Finally, using the Clausius-Mossoti relation,  we provide estimates of 
the dielectric constant of crystalline solids of these fullerenes.
Assuming that like C$_{60}$, all larger fullerenes form an FCC crystal, the number density
of fullerene is then 4/V, where V=$(\sqrt(8)r)^3$. The fullerene radius used 
here is the effective radius as used above (to include the $\pi$-electron spill-out effect).
Using a value of $\delta = 1.23$\AA\, results in 
a smaller lattice constant than the experimental value (14.17 \AA)\cite{C60_lattice_constant}
but gives a
dielectric constant of 4.50, which is in good agreement with experimental
values (4.0 - 4.5)\cite{hebard:2109}. 
The dielectric constants calculated for the larger fullerenes
using the same $\delta$ value  are in the range 4.7-5.6 (See last column in Table \ref{table:pol}).
These values may vary slightly with the $\delta$ but the trend should be the same.

\section{Conclusions}

    In summary, static electric response of the icosahedral 
C$_{60}$, C$_{180}$, C$_{240}$, C$_{540}$,
C$_{720}$, C$_{960}$, C$_{1500}$, and C$_{2160}$ fullerenes 
is studied in detail by an all-electron 
first principles density-functional methods. Quantitative estimates of 
the dipole polarizability determined herein provide better understanding
of electronic response of fullerenes amongst earlier inconsistent predictions 
from various models.  This work shows that quantum size effects in 
polarizability are substantially quenched by C$_{2160}$ in sharp contrast to 
previous tight-binding predictions. These calculations also signals the beginning of 
an era where tens of thousands of basis functions will be used for accurate
electronic structure calculations.

\acknowledgments 

        This work is supported in part by 
the National Science Foundation through CREST grant,
by the University of Texas at El Paso (UTEP startup funds) and  partly
the Office of Naval Research. Authors acknowledge the computer time 
at the UTEP Cray acquired using ONR 05PR07548-00 grant.

\begin{table}
\caption{The details of the Gaussian basis set used in polarizability calculation.
The exponent ($\alpha$), contraction coefficients c(1s), c(2s), and 
c(2p) are listed below.  The basis set also contains single Gaussians. 
For the s and p functions we use $\alpha_{10}-\alpha_{12}$. 
The d functions use  $\alpha_{9}-\alpha_{12}$. 
\label{tab:basis}
}
\begin{tabular}{rcccccccc}
\hline
 $\alpha$(Bohr$^{-2}$) &   c(1s)   &   c(2s)  &   c(2p)  \\
\hline
  22213.361000  &         0.197922 &        -0.045005 &         0.023139 \\
    3331.737000  &         0.369990 &        -0.084621 &         0.042649 \\
     757.901350  &         0.636446 &        -0.144966 &         0.074659 \\
     214.543720  &         1.012493 &        -0.235356 &         0.120241 \\
      69.924889  &         1.448079 &        -0.342154 &         0.183512 \\
      25.086135  &         1.717369 &        -0.445951 &         0.247068 \\
       9.591042  &         1.493193 &        -0.452640 &         0.307142 \\
       3.802456  &         0.689872 &        -0.322164 &         0.313727 \\
       1.489185  &         0.086072 &        -0.012988 &         0.267263 \\
       0.574877  &        -0.001657 &         0.201355 &         0.147566 \\
       0.214947  &         0.000378 &         0.127699 &         0.047586 \\
       0.077210  &        -0.000047 &         0.014135 &         0.007280 \\
\hline
\end{tabular}
\end{table}

\begin{figure}
\epsfig{file=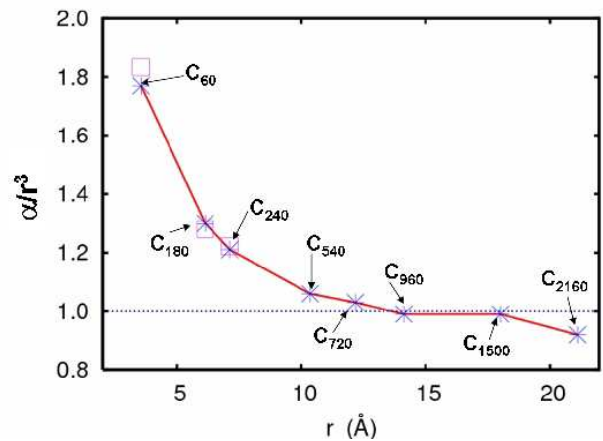,width=8.5cm,clip=true}
\caption{ (Color online) The ratio of 
polarizability, $\alpha_{RPA}$, and volume, $\overline{r}^3$,
as a function of fullerene radius $\overline{r}$. The squares 
represent polarizability obtained by the finite-field method.
 \label{fig:fpol}
}
\end{figure}

\begin{thebibliography}{50}
\expandafter\ifx\csname natexlab\endcsname\relax\def\natexlab#1{#1}\fi
\expandafter\ifx\csname bibnamefont\endcsname\relax
  \def\bibnamefont#1{#1}\fi
\expandafter\ifx\csname bibfnamefont\endcsname\relax
  \def\bibfnamefont#1{#1}\fi
\expandafter\ifx\csname citenamefont\endcsname\relax
  \def\citenamefont#1{#1}\fi
\expandafter\ifx\csname url\endcsname\relax
  \def\url#1{\texttt{#1}}\fi
\expandafter\ifx\csname urlprefix\endcsname\relax\def\urlprefix{URL }\fi
\providecommand{\bibinfo}[2]{#2}
\providecommand{\eprint}[2][]{\url{#2}}

\bibitem[{\citenamefont{Haberland et~al.}(2005)\citenamefont{Haberland,
  Hippler, Donges, Kostko, Schmidt, and von Issendorff}}]{haberland:035701}
\bibinfo{author}{\bibfnamefont{H.}~\bibnamefont{Haberland}},
  \bibinfo{author}{\bibfnamefont{T.}~\bibnamefont{Hippler}},
  \bibinfo{author}{\bibfnamefont{J.}~\bibnamefont{Donges}},
  \bibinfo{author}{\bibfnamefont{O.}~\bibnamefont{Kostko}},
  \bibinfo{author}{\bibfnamefont{M.}~\bibnamefont{Schmidt}}, \bibnamefont{and}
  \bibinfo{author}{\bibfnamefont{B.}~\bibnamefont{von Issendorff}},
  \bibinfo{journal}{Physical Review Letters} \textbf{\bibinfo{volume}{94}},
  \bibinfo{eid}{035701} (pages~\bibinfo{numpages}{4}) (\bibinfo{year}{2005}).

\bibitem[{\citenamefont{Ramos et~al.}(2005)\citenamefont{Ramos,
  Furthm\"{u}ller, and Bechstedt}}]{ramos:045351}
\bibinfo{author}{\bibfnamefont{L.~E.} \bibnamefont{Ramos}},
  \bibinfo{author}{\bibfnamefont{J.}~\bibnamefont{Furthm\"{u}ller}},
  \bibnamefont{and}
  \bibinfo{author}{\bibfnamefont{F.}~\bibnamefont{Bechstedt}},
  \bibinfo{journal}{Physical Review B (Condensed Matter and Materials Physics)}
  \textbf{\bibinfo{volume}{72}}, \bibinfo{eid}{045351}
  (pages~\bibinfo{numpages}{8}) (\bibinfo{year}{2005}).

\bibitem[{\citenamefont{Wang et~al.}(2007)\citenamefont{Wang, Zhang, Lee,
  Niehaus, and Frauenheim}}]{wang:123116}
\bibinfo{author}{\bibfnamefont{X.}~\bibnamefont{Wang}},
  \bibinfo{author}{\bibfnamefont{R.~Q.} \bibnamefont{Zhang}},
  \bibinfo{author}{\bibfnamefont{S.~T.} \bibnamefont{Lee}},
  \bibinfo{author}{\bibfnamefont{T.~A.} \bibnamefont{Niehaus}},
  \bibnamefont{and}
  \bibinfo{author}{\bibfnamefont{T.}~\bibnamefont{Frauenheim}},
  \bibinfo{journal}{Applied Physics Letters} \textbf{\bibinfo{volume}{90}},
  \bibinfo{eid}{123116} (pages~\bibinfo{numpages}{3}) (\bibinfo{year}{2007}),
  \urlprefix\url{http://link.aip.org/link/?APL/90/123116/1}.

\bibitem[{\citenamefont{Yang et~al.}(2007)\citenamefont{Yang, Spataru, Louie,
  and Chou}}]{yang:201304}
\bibinfo{author}{\bibfnamefont{L.}~\bibnamefont{Yang}},
  \bibinfo{author}{\bibfnamefont{C.~D.} \bibnamefont{Spataru}},
  \bibinfo{author}{\bibfnamefont{S.~G.} \bibnamefont{Louie}}, \bibnamefont{and}
  \bibinfo{author}{\bibfnamefont{M.~Y.} \bibnamefont{Chou}},
  \bibinfo{journal}{Physical Review B (Condensed Matter and Materials Physics)}
  \textbf{\bibinfo{volume}{75}}, \bibinfo{eid}{201304}
  (pages~\bibinfo{numpages}{4}) (\bibinfo{year}{2007}).

\bibitem[{\citenamefont{Raty et~al.}(2003)\citenamefont{Raty, Galli, Bostedt,
  van Buuren, and Terminello}}]{raty:037401}
\bibinfo{author}{\bibfnamefont{J.-Y.} \bibnamefont{Raty}},
  \bibinfo{author}{\bibfnamefont{G.}~\bibnamefont{Galli}},
  \bibinfo{author}{\bibfnamefont{C.}~\bibnamefont{Bostedt}},
  \bibinfo{author}{\bibfnamefont{T.~W.} \bibnamefont{van Buuren}},
  \bibnamefont{and} \bibinfo{author}{\bibfnamefont{L.~J.}
  \bibnamefont{Terminello}}, \bibinfo{journal}{Physical Review Letters}
  \textbf{\bibinfo{volume}{90}}, \bibinfo{eid}{037401}
  (pages~\bibinfo{numpages}{4}) (\bibinfo{year}{2003}),
  \urlprefix\url{http://link.aps.org/abstract/PRL/v90/e037401}.

\bibitem[{\citenamefont{Drummond et~al.}(2005)\citenamefont{Drummond,
  Williamson, Needs, and Galli}}]{drummond:096801}
\bibinfo{author}{\bibfnamefont{N.~D.} \bibnamefont{Drummond}},
  \bibinfo{author}{\bibfnamefont{A.~J.} \bibnamefont{Williamson}},
  \bibinfo{author}{\bibfnamefont{R.~J.} \bibnamefont{Needs}}, \bibnamefont{and}
  \bibinfo{author}{\bibfnamefont{G.}~\bibnamefont{Galli}},
  \bibinfo{journal}{Physical Review Letters} \textbf{\bibinfo{volume}{95}},
  \bibinfo{eid}{096801} (pages~\bibinfo{numpages}{4}) (\bibinfo{year}{2005}),
  \urlprefix\url{http://link.aps.org/abstract/PRL/v95/e096801}.

\bibitem[{\citenamefont{{Volokitin} et~al.}(1996)\citenamefont{{Volokitin},
  {Sinzig}, {de Jongh}, {Schmid}, {Vargaftik}, and
  {Moiseevi}}}]{1996Natur.384..621V}
\bibinfo{author}{\bibfnamefont{Y.}~\bibnamefont{{Volokitin}}},
  \bibinfo{author}{\bibfnamefont{J.}~\bibnamefont{{Sinzig}}},
  \bibinfo{author}{\bibfnamefont{L.~J.} \bibnamefont{{de Jongh}}},
  \bibinfo{author}{\bibfnamefont{G.}~\bibnamefont{{Schmid}}},
  \bibinfo{author}{\bibfnamefont{M.~N.} \bibnamefont{{Vargaftik}}},
  \bibnamefont{and} \bibinfo{author}{\bibfnamefont{I.~I.}
  \bibnamefont{{Moiseevi}}}, \bibinfo{journal}{\nat}
  \textbf{\bibinfo{volume}{384}}, \bibinfo{pages}{621} (\bibinfo{year}{1996}).

\bibitem[{\citenamefont{Breaux et~al.}(2005)\citenamefont{Breaux, Neal, Cao,
  and Jarrold}}]{breaux:173401}
\bibinfo{author}{\bibfnamefont{G.~A.} \bibnamefont{Breaux}},
  \bibinfo{author}{\bibfnamefont{C.~M.} \bibnamefont{Neal}},
  \bibinfo{author}{\bibfnamefont{B.}~\bibnamefont{Cao}}, \bibnamefont{and}
  \bibinfo{author}{\bibfnamefont{M.~F.} \bibnamefont{Jarrold}},
  \bibinfo{journal}{Physical Review Letters} \textbf{\bibinfo{volume}{94}},
  \bibinfo{eid}{173401} (pages~\bibinfo{numpages}{4}) (\bibinfo{year}{2005}),
  \urlprefix\url{http://link.aps.org/abstract/PRL/v94/e173401}.

\bibitem[{\citenamefont{Vasiliev et~al.}(2001)\citenamefont{Vasiliev, Ogut, and
  Chelikowsky}}]{vasiliev:1813}
\bibinfo{author}{\bibfnamefont{I.}~\bibnamefont{Vasiliev}},
  \bibinfo{author}{\bibfnamefont{S.}~\bibnamefont{Ogut}}, \bibnamefont{and}
  \bibinfo{author}{\bibfnamefont{J.~R.} \bibnamefont{Chelikowsky}},
  \bibinfo{journal}{Physical Review Letters} \textbf{\bibinfo{volume}{86}},
  \bibinfo{pages}{1813} (\bibinfo{year}{2001}),
  \urlprefix\url{http://link.aps.org/abstract/PRL/v86/p1813}.

\bibitem[{\citenamefont{Gueorguiev et~al.}(2004)\citenamefont{Gueorguiev,
  Pacheco, and Tomanek}}]{Pacheco}
\bibinfo{author}{\bibfnamefont{G.~K.} \bibnamefont{Gueorguiev}},
  \bibinfo{author}{\bibfnamefont{J.~M.} \bibnamefont{Pacheco}},
  \bibnamefont{and} \bibinfo{author}{\bibfnamefont{D.}~\bibnamefont{Tomanek}},
  \bibinfo{journal}{Phys. Rev. Lett.} \textbf{\bibinfo{volume}{92}},
  \bibinfo{eid}{215501} (pages~\bibinfo{numpages}{4}) (\bibinfo{year}{2004}).

\bibitem[{\citenamefont{Ruiz et~al.}(2001)\citenamefont{Ruiz, Breton, and
  Llorente}}]{C70:376}
\bibinfo{author}{\bibfnamefont{A.}~\bibnamefont{Ruiz}},
  \bibinfo{author}{\bibfnamefont{J.}~\bibnamefont{Breton}}, \bibnamefont{and}
  \bibinfo{author}{\bibfnamefont{J.~M.~G.} \bibnamefont{Llorente}},
  \bibinfo{journal}{Journal of Chemical Physics}
  \textbf{\bibinfo{volume}{114}}, \bibinfo{pages}{1272} (\bibinfo{year}{2001}).

\bibitem[{\citenamefont{Hu and Ruckenstein}(2005)}]{hu:214708}
\bibinfo{author}{\bibfnamefont{Y.~H.} \bibnamefont{Hu}} \bibnamefont{and}
  \bibinfo{author}{\bibfnamefont{E.}~\bibnamefont{Ruckenstein}},
  \bibinfo{journal}{The Journal of Chemical Physics}
  \textbf{\bibinfo{volume}{123}}, \bibinfo{eid}{214708}
  (pages~\bibinfo{numpages}{4}) (\bibinfo{year}{2005}).

\bibitem[{\citenamefont{I.Groth}(2007)}]{Astro}
\bibinfo{author}{\bibfnamefont{S.}~\bibnamefont{I.Groth}},
  \bibinfo{journal}{The Astrophysical Journal} \textbf{\bibinfo{volume}{661}},
  \bibinfo{pages}{L167} (\bibinfo{year}{2007}).

\bibitem[{\citenamefont{Werpetinski and Cook}(1995)}]{PhysRevA.52.R3397}
\bibinfo{author}{\bibfnamefont{K.~S.} \bibnamefont{Werpetinski}}
  \bibnamefont{and} \bibinfo{author}{\bibfnamefont{M.}~\bibnamefont{Cook}},
  \bibinfo{journal}{Phys. Rev. A} \textbf{\bibinfo{volume}{52}},
  \bibinfo{pages}{R3397} (\bibinfo{year}{1995}).

\bibitem[{\citenamefont{Werpetinski and Cook}(1997)}]{werpetinski:7124}
\bibinfo{author}{\bibfnamefont{K.~S.} \bibnamefont{Werpetinski}}
  \bibnamefont{and} \bibinfo{author}{\bibfnamefont{M.}~\bibnamefont{Cook}},
  \bibinfo{journal}{The Journal of Chemical Physics}
  \textbf{\bibinfo{volume}{106}}, \bibinfo{pages}{7124} (\bibinfo{year}{1997}),
  \urlprefix\url{http://link.aip.org/link/?JCP/106/7124/1}.

\bibitem[{\citenamefont{Dunlap}(2003)}]{RF:204}
\bibinfo{author}{\bibfnamefont{B.~I.} \bibnamefont{Dunlap}},
  \bibinfo{journal}{Journal of Physical Chemistry a}
  \textbf{\bibinfo{volume}{107}}, \bibinfo{pages}{10082}
  (\bibinfo{year}{2003}), \bibinfo{note}{pT: J}.

\bibitem[{\citenamefont{Zope and Dunlap}(2005{\natexlab{a}})}]{RF:110}
\bibinfo{author}{\bibfnamefont{R.~R.} \bibnamefont{Zope}} \bibnamefont{and}
  \bibinfo{author}{\bibfnamefont{B.~I.} \bibnamefont{Dunlap}},
  \bibinfo{journal}{Physical Review B} \textbf{\bibinfo{volume}{71}},
  \bibinfo{pages}{193104} (\bibinfo{year}{2005}{\natexlab{a}}),
  \bibinfo{note}{pT: J}.

\bibitem[{\citenamefont{Zope and Dunlap}(2005{\natexlab{b}})}]{RF:108}
\bibinfo{author}{\bibfnamefont{R.~R.} \bibnamefont{Zope}} \bibnamefont{and}
  \bibinfo{author}{\bibfnamefont{B.~I.} \bibnamefont{Dunlap}},
  \bibinfo{journal}{Journal of Chemical Theory and Computation}
  \textbf{\bibinfo{volume}{1}}, \bibinfo{pages}{1193}
  (\bibinfo{year}{2005}{\natexlab{b}}), \bibinfo{note}{pT: J}.

\bibitem[{\citenamefont{Zope and Dunlap}(2006)}]{RF:194}
\bibinfo{author}{\bibfnamefont{R.~R.} \bibnamefont{Zope}} \bibnamefont{and}
  \bibinfo{author}{\bibfnamefont{B.~I.} \bibnamefont{Dunlap}},
  \bibinfo{journal}{Journal of Chemical Physics}
  \textbf{\bibinfo{volume}{124}}, \bibinfo{pages}{044107}
  (\bibinfo{year}{2006}), \bibinfo{note}{pT: J}.

\bibitem[{\citenamefont{Zope and Dunlap}(2005{\natexlab{c}})}]{RF:3}
\bibinfo{author}{\bibfnamefont{R.~R.} \bibnamefont{Zope}} \bibnamefont{and}
  \bibinfo{author}{\bibfnamefont{B.~I.} \bibnamefont{Dunlap}},
  \bibinfo{journal}{Physical Review B} \textbf{\bibinfo{volume}{72}},
  \bibinfo{pages}{45439} (\bibinfo{year}{2005}{\natexlab{c}}).

\bibitem[{\citenamefont{Zope et~al.}(2004)\citenamefont{Zope, Baruah, Pederson,
  and Dunlap}}]{RF:4}
\bibinfo{author}{\bibfnamefont{R.~R.} \bibnamefont{Zope}},
  \bibinfo{author}{\bibfnamefont{T.}~\bibnamefont{Baruah}},
  \bibinfo{author}{\bibfnamefont{M.~R.} \bibnamefont{Pederson}},
  \bibnamefont{and} \bibinfo{author}{\bibfnamefont{B.~I.}
  \bibnamefont{Dunlap}}, \bibinfo{journal}{Arxiv preprint physics/0407031}
  (\bibinfo{year}{2004}).

\bibitem[{\citenamefont{Zope and Dunlap}(2004)}]{RF:119}
\bibinfo{author}{\bibfnamefont{R.~R.} \bibnamefont{Zope}} \bibnamefont{and}
  \bibinfo{author}{\bibfnamefont{B.~I.} \bibnamefont{Dunlap}},
  \bibinfo{journal}{Chemical Physics Letters} \textbf{\bibinfo{volume}{386}},
  \bibinfo{pages}{403} (\bibinfo{year}{2004}), \bibinfo{note}{pT: J}.

\bibitem[{\citenamefont{McLean and Chandler}(1980)}]{mclean:5639}
\bibinfo{author}{\bibfnamefont{A.~D.} \bibnamefont{McLean}} \bibnamefont{and}
  \bibinfo{author}{\bibfnamefont{G.~S.} \bibnamefont{Chandler}},
  \bibinfo{journal}{The Journal of Chemical Physics}
  \textbf{\bibinfo{volume}{72}}, \bibinfo{pages}{5639} (\bibinfo{year}{1980}),
  \urlprefix\url{http://link.aip.org/link/?JCP/72/5639/1}.

\bibitem[{\citenamefont{Dunlap and Zope}(2006)}]{RF:106}
\bibinfo{author}{\bibfnamefont{B.~I.} \bibnamefont{Dunlap}} \bibnamefont{and}
  \bibinfo{author}{\bibfnamefont{R.~R.} \bibnamefont{Zope}},
  \bibinfo{journal}{Chemical Physics Letters} \textbf{\bibinfo{volume}{422}},
  \bibinfo{pages}{451} (\bibinfo{year}{2006}), \bibinfo{note}{pT: J}.

\bibitem[{\citenamefont{Perdew et~al.}(1996)\citenamefont{Perdew, Burke, and
  Ernzerhof}}]{RF:183}
\bibinfo{author}{\bibfnamefont{J.~P.} \bibnamefont{Perdew}},
  \bibinfo{author}{\bibfnamefont{K.}~\bibnamefont{Burke}}, \bibnamefont{and}
  \bibinfo{author}{\bibfnamefont{M.}~\bibnamefont{Ernzerhof}},
  \bibinfo{journal}{Physical Review Letters} \textbf{\bibinfo{volume}{77}},
  \bibinfo{pages}{3865} (\bibinfo{year}{1996}), \bibinfo{note}{pT: J}.

\bibitem[{\citenamefont{Pederson et~al.}(2005)\citenamefont{Pederson, Baruah,
  Allen, and Schmidt}}]{PBAS05}
\bibinfo{author}{\bibfnamefont{M.}~\bibnamefont{Pederson}},
  \bibinfo{author}{\bibfnamefont{T.}~\bibnamefont{Baruah}},
  \bibinfo{author}{\bibfnamefont{P.}~\bibnamefont{Allen}}, \bibnamefont{and}
  \bibinfo{author}{\bibfnamefont{C.}~\bibnamefont{Schmidt}},
  \bibinfo{journal}{Journal of Chemical Theory and Computation}
  \textbf{\bibinfo{volume}{1}}, \bibinfo{pages}{590} (\bibinfo{year}{2005}),
  ISSN \bibinfo{issn}{1549-9618}.

\bibitem[{\citenamefont{Zope}(2007)}]{RRZ07}
\bibinfo{author}{\bibfnamefont{R.~R.} \bibnamefont{Zope}},
  \bibinfo{journal}{Journal of Physics B: Atomic, Molecular and Optical
  Physics} \textbf{\bibinfo{volume}{40}}, \bibinfo{pages}{3491}
  (\bibinfo{year}{2007}),
  \urlprefix\url{http://stacks.iop.org/0953-4075/40/3491}.

\bibitem[{\citenamefont{Cohen and Roothaan}(1965)}]{cohen:s34}
\bibinfo{author}{\bibfnamefont{H.~D.} \bibnamefont{Cohen}} \bibnamefont{and}
  \bibinfo{author}{\bibfnamefont{C.~C.~J.} \bibnamefont{Roothaan}},
  \bibinfo{journal}{The Journal of Chemical Physics}
  \textbf{\bibinfo{volume}{43}}, \bibinfo{pages}{S34} (\bibinfo{year}{1965}).

\bibitem[{\citenamefont{Bartlett and Purvis}(1979)}]{Bartlett_FF}
\bibinfo{author}{\bibfnamefont{R.~J.} \bibnamefont{Bartlett}} \bibnamefont{and}
  \bibinfo{author}{\bibfnamefont{G.~D.} \bibnamefont{Purvis}},
  \bibinfo{journal}{Phys. Rev. A} \textbf{\bibinfo{volume}{20}},
  \bibinfo{pages}{1313} (\bibinfo{year}{1979}).

\bibitem[{\citenamefont{Kurtz et~al.}(1982)\citenamefont{Kurtz, Stewart, and
  Dieter}}]{Kurtz82}
\bibinfo{author}{\bibfnamefont{H.~A.} \bibnamefont{Kurtz}},
  \bibinfo{author}{\bibfnamefont{J.~J.~P.} \bibnamefont{Stewart}},
  \bibnamefont{and} \bibinfo{author}{\bibfnamefont{K.~M.}
  \bibnamefont{Dieter}}, \bibinfo{journal}{J. Comput. Chem.}
  \textbf{\bibinfo{volume}{11}} (\bibinfo{year}{1982}).

\bibitem[{\citenamefont{Guan et~al.}(1995)\citenamefont{Guan, Casida, K\"oster,
  and Salahub}}]{Salahub_Pol}
\bibinfo{author}{\bibfnamefont{J.}~\bibnamefont{Guan}},
  \bibinfo{author}{\bibfnamefont{M.~E.} \bibnamefont{Casida}},
  \bibinfo{author}{\bibfnamefont{A.~M.} \bibnamefont{K\"oster}},
  \bibnamefont{and} \bibinfo{author}{\bibfnamefont{D.~R.}
  \bibnamefont{Salahub}}, \bibinfo{journal}{Phys. Rev. B}
  \textbf{\bibinfo{volume}{52}}, \bibinfo{pages}{2184} (\bibinfo{year}{1995}).

\bibitem[{\citenamefont{Blundell et~al.}(2000)\citenamefont{Blundell, Guet, and
  Zope}}]{Na_Pol}
\bibinfo{author}{\bibfnamefont{S.~A.} \bibnamefont{Blundell}},
  \bibinfo{author}{\bibfnamefont{C.}~\bibnamefont{Guet}}, \bibnamefont{and}
  \bibinfo{author}{\bibfnamefont{R.~R.} \bibnamefont{Zope}},
  \bibinfo{journal}{Phys. Rev. Lett.} \textbf{\bibinfo{volume}{84}},
  \bibinfo{pages}{4826} (\bibinfo{year}{2000}).

\bibitem[{\citenamefont{Jackson et~al.}(1999)\citenamefont{Jackson, Pederson,
  Wang, and Ho}}]{Jackson}
\bibinfo{author}{\bibfnamefont{K.}~\bibnamefont{Jackson}},
  \bibinfo{author}{\bibfnamefont{M.}~\bibnamefont{Pederson}},
  \bibinfo{author}{\bibfnamefont{C.-Z.} \bibnamefont{Wang}}, \bibnamefont{and}
  \bibinfo{author}{\bibfnamefont{K.-M.} \bibnamefont{Ho}},
  \bibinfo{journal}{Phys. Rev. A} \textbf{\bibinfo{volume}{59}},
  \bibinfo{pages}{3685} (\bibinfo{year}{1999}).

\bibitem[{\citenamefont{Pederson and Jackson}(1991)}]{RF:69}
\bibinfo{author}{\bibfnamefont{M.~R.} \bibnamefont{Pederson}} \bibnamefont{and}
  \bibinfo{author}{\bibfnamefont{K.~A.} \bibnamefont{Jackson}},
  \bibinfo{journal}{PHYSICAL REVIEW B.CONDENSED MATTER}
  \textbf{\bibinfo{volume}{43}}, \bibinfo{pages}{7312} (\bibinfo{year}{1991}),
  \bibinfo{note}{pUBM: Print; ppublish}.

\bibitem[{\citenamefont{Pederson and Jackson}(1990)}]{RF:74}
\bibinfo{author}{\bibfnamefont{M.~R.} \bibnamefont{Pederson}} \bibnamefont{and}
  \bibinfo{author}{\bibfnamefont{K.~A.} \bibnamefont{Jackson}},
  \bibinfo{journal}{PHYSICAL REVIEW B.CONDENSED MATTER}
  \textbf{\bibinfo{volume}{41}}, \bibinfo{pages}{7453} (\bibinfo{year}{1990}),
  \bibinfo{note}{pUBM: Print; ppublish}.

\bibitem[{\citenamefont{Jackson and Pederson}(1990{\natexlab{a}})}]{RF:71}
\bibinfo{author}{\bibfnamefont{K.}~\bibnamefont{Jackson}} \bibnamefont{and}
  \bibinfo{author}{\bibfnamefont{M.~R.} \bibnamefont{Pederson}},
  \bibinfo{journal}{PHYSICAL REVIEW B.CONDENSED MATTER}
  \textbf{\bibinfo{volume}{42}}, \bibinfo{pages}{3276}
  (\bibinfo{year}{1990}{\natexlab{a}}), \bibinfo{note}{pUBM: Print; ppublish}.

\bibitem[{\citenamefont{Jackson and Pederson}(1990{\natexlab{b}})}]{RF:184}
\bibinfo{author}{\bibfnamefont{K.}~\bibnamefont{Jackson}} \bibnamefont{and}
  \bibinfo{author}{\bibfnamefont{M.~R.} \bibnamefont{Pederson}},
  \bibinfo{journal}{Physical Review B} \textbf{\bibinfo{volume}{42}},
  \bibinfo{pages}{3276} (\bibinfo{year}{1990}{\natexlab{b}}),
  \bibinfo{note}{pT: J}.

\bibitem[{\citenamefont{Porezag and Pederson}(1999)}]{RF:181}
\bibinfo{author}{\bibfnamefont{D.}~\bibnamefont{Porezag}} \bibnamefont{and}
  \bibinfo{author}{\bibfnamefont{M.~R.} \bibnamefont{Pederson}},
  \bibinfo{journal}{Physical Review a} \textbf{\bibinfo{volume}{60}},
  \bibinfo{pages}{2840} (\bibinfo{year}{1999}), \bibinfo{note}{pT: J}.

\bibitem[{\citenamefont{Pederson and Baruah}(2005)}]{PB_Pol}
\bibinfo{author}{\bibfnamefont{M.~R.} \bibnamefont{Pederson}} \bibnamefont{and}
  \bibinfo{author}{\bibfnamefont{T.}~\bibnamefont{Baruah}},
  \bibinfo{journal}{Lecture Series in Computer and Computational Sciences}
  \textbf{\bibinfo{volume}{3}}, \bibinfo{pages}{156} (\bibinfo{year}{2005}).

\bibitem[{\citenamefont{Westin et~al.}(1996)\citenamefont{Westin, Ros\'{e}n,
  Velde, and Baerends}}]{Westin}
\bibinfo{author}{\bibfnamefont{E.}~\bibnamefont{Westin}},
  \bibinfo{author}{\bibfnamefont{A.}~\bibnamefont{Ros\'{e}n}},
  \bibinfo{author}{\bibfnamefont{G.~T.} \bibnamefont{Velde}}, \bibnamefont{and}
  \bibinfo{author}{\bibfnamefont{E.~J.} \bibnamefont{Baerends}},
  \bibinfo{journal}{Journal of Physics B: Atomic, Molecular and Optical
  Physics} \textbf{\bibinfo{volume}{29}}, \bibinfo{pages}{5087}
  (\bibinfo{year}{1996}).

\bibitem[{\citenamefont{Pederson and Quong}(1992)}]{RF:59}
\bibinfo{author}{\bibfnamefont{M.~R.} \bibnamefont{Pederson}} \bibnamefont{and}
  \bibinfo{author}{\bibfnamefont{A.~A.} \bibnamefont{Quong}},
  \bibinfo{journal}{Phys. Rev. B} \textbf{\bibinfo{volume}{46}},
  \bibinfo{pages}{13584} (\bibinfo{year}{1992}).

\bibitem[{\citenamefont{Zope et~al.}(2007)\citenamefont{Zope, Baruah, Pederson,
  and Dunlap}}]{Zope:cond-mat0701466}
\bibinfo{author}{\bibfnamefont{R.~R.} \bibnamefont{Zope}},
  \bibinfo{author}{\bibfnamefont{T.}~\bibnamefont{Baruah}},
  \bibinfo{author}{\bibfnamefont{M.~R.} \bibnamefont{Pederson}},
  \bibnamefont{and} \bibinfo{author}{\bibfnamefont{B.~I.} \bibnamefont{Dunlap}}
  (\bibinfo{year}{2007}), \eprint{cond-mat/0701466}.

\bibitem[{\citenamefont{Benedict et~al.}(1995)\citenamefont{Benedict, Louie,
  and Cohen}}]{Benedict}
\bibinfo{author}{\bibfnamefont{L.~X.} \bibnamefont{Benedict}},
  \bibinfo{author}{\bibfnamefont{S.~G.} \bibnamefont{Louie}}, \bibnamefont{and}
  \bibinfo{author}{\bibfnamefont{M.~L.} \bibnamefont{Cohen}},
  \bibinfo{journal}{Phys. Rev. B} \textbf{\bibinfo{volume}{52}},
  \bibinfo{pages}{8541} (\bibinfo{year}{1995}).

\bibitem[{\citenamefont{Bertsch et~al.}(1991)\citenamefont{Bertsch, Bulgac,
  Tom\'anek, and Wang}}]{Bertsch}
\bibinfo{author}{\bibfnamefont{G.~F.} \bibnamefont{Bertsch}},
  \bibinfo{author}{\bibfnamefont{A.}~\bibnamefont{Bulgac}},
  \bibinfo{author}{\bibfnamefont{D.}~\bibnamefont{Tom\'anek}},
  \bibnamefont{and} \bibinfo{author}{\bibfnamefont{Y.}~\bibnamefont{Wang}},
  \bibinfo{journal}{Phys. Rev. Lett.} \textbf{\bibinfo{volume}{67}},
  \bibinfo{pages}{2690} (\bibinfo{year}{1991}).

\bibitem[{\citenamefont{Ruiz et~al.}(2005)\citenamefont{Ruiz, Breton, and
  Llorente}}]{Ruiz}
\bibinfo{author}{\bibfnamefont{A.}~\bibnamefont{Ruiz}},
  \bibinfo{author}{\bibfnamefont{J.}~\bibnamefont{Breton}}, \bibnamefont{and}
  \bibinfo{author}{\bibfnamefont{J.~M.~G.} \bibnamefont{Llorente}},
  \bibinfo{journal}{Physical Review Letters} \textbf{\bibinfo{volume}{94}},
  \bibinfo{eid}{105501} (pages~\bibinfo{numpages}{4}) (\bibinfo{year}{2005}).

\bibitem[{\citenamefont{Brown et~al.}(2006)\citenamefont{Brown, Cao, Musfeldt,
  Dragoe, Cimpoesu, Ito, Takagi, and Cross}}]{Musfeldt}
\bibinfo{author}{\bibfnamefont{S.}~\bibnamefont{Brown}},
  \bibinfo{author}{\bibfnamefont{J.}~\bibnamefont{Cao}},
  \bibinfo{author}{\bibfnamefont{J.~L.} \bibnamefont{Musfeldt}},
  \bibinfo{author}{\bibfnamefont{N.}~\bibnamefont{Dragoe}},
  \bibinfo{author}{\bibfnamefont{F.}~\bibnamefont{Cimpoesu}},
  \bibinfo{author}{\bibfnamefont{S.}~\bibnamefont{Ito}},
  \bibinfo{author}{\bibfnamefont{H.}~\bibnamefont{Takagi}}, \bibnamefont{and}
  \bibinfo{author}{\bibfnamefont{R.~J.} \bibnamefont{Cross}},
  \bibinfo{journal}{Physical Review B (Condensed Matter and Materials Physics)}
  \textbf{\bibinfo{volume}{73}}, \bibinfo{eid}{125446}
  (pages~\bibinfo{numpages}{6}) (\bibinfo{year}{2006}).

\bibitem[{\citenamefont{Schelling and Keblinski}(2003)}]{PhysRevB.68.035425}
\bibinfo{author}{\bibfnamefont{P.~K.} \bibnamefont{Schelling}}
  \bibnamefont{and}
  \bibinfo{author}{\bibfnamefont{P.}~\bibnamefont{Keblinski}},
  \bibinfo{journal}{Phys. Rev. B} \textbf{\bibinfo{volume}{68}},
  \bibinfo{pages}{035425} (\bibinfo{year}{2003}).

\bibitem[{\citenamefont{Yoshida et~al.}(2007)\citenamefont{Yoshida, Kurui,
  Oshima, and Takayanagi}}]{Japan}
\bibinfo{author}{\bibfnamefont{M.}~\bibnamefont{Yoshida}},
  \bibinfo{author}{\bibfnamefont{Y.}~\bibnamefont{Kurui}},
  \bibinfo{author}{\bibfnamefont{Y.}~\bibnamefont{Oshima}}, \bibnamefont{and}
  \bibinfo{author}{\bibfnamefont{K.}~\bibnamefont{Takayanagi}},
  \bibinfo{journal}{Jap. J. Appl. Phys.} \textbf{\bibinfo{volume}{46}},
  \bibinfo{pages}{L67} (\bibinfo{year}{2007}).

\bibitem[{\citenamefont{Eklund et~al.}(1995)\citenamefont{Eklund, Rao, and
  et~al.}}]{C60_lattice_constant}
\bibinfo{author}{\bibfnamefont{P.~C.} \bibnamefont{Eklund}},
  \bibinfo{author}{\bibfnamefont{A.~M.} \bibnamefont{Rao}}, \bibnamefont{and}
  \bibinfo{author}{\bibfnamefont{Y.~W.} \bibnamefont{et~al.}},
  \bibinfo{journal}{Thin Solid Films} \textbf{\bibinfo{volume}{257}},
  \bibinfo{pages}{7590} (\bibinfo{year}{1995}).

\bibitem[{\citenamefont{Hebard et~al.}(1991)\citenamefont{Hebard, Haddon,
  Fleming, and Kortan}}]{hebard:2109}
\bibinfo{author}{\bibfnamefont{A.~F.} \bibnamefont{Hebard}},
  \bibinfo{author}{\bibfnamefont{R.~C.} \bibnamefont{Haddon}},
  \bibinfo{author}{\bibfnamefont{R.~M.} \bibnamefont{Fleming}},
  \bibnamefont{and} \bibinfo{author}{\bibfnamefont{A.~R.}
  \bibnamefont{Kortan}}, \bibinfo{journal}{Applied Physics Letters}
  \textbf{\bibinfo{volume}{59}}, \bibinfo{pages}{2109} (\bibinfo{year}{1991}).

\end{thebibliography}

\end{document}